\documentclass[lettersize,journal]{IEEEtran}
\usepackage{amsmath,amsfonts}
 \usepackage{bm}
\usepackage{subfigure}
\usepackage{algorithmic}
\usepackage{array}
\usepackage[caption=false,font=normalsize,labelfont=sf,textfont=sf]{subfig}
\usepackage{textcomp}
\usepackage{stfloats}
\usepackage{url}
\usepackage{verbatim}
\usepackage{graphicx}
\def\BibTeX{{\rm B\kern-.05em{\sc i\kern-.025em b}\kern-.08em
    T\kern-.1667em\lower.7ex\hbox{E}\kern-.125emX}}
\usepackage{balance}
\usepackage{color}
\begin{document}
\title{ A General Sensing-assisted Channel Estimation Framework in Distributed MIMO Network}
\author{ Hui Zhou ~\IEEEmembership{Member,~IEEE,} Xiaolan Liu ~\IEEEmembership{Member,~IEEE,} Sangarapillai Lambotharan ~\IEEEmembership{Senior Member,~IEEE,
\thanks{Hui Zhou is with Centre for Future Transport and Cities, Coventry University, U.K. (Email:hui.zhou@coventry.ac.uk).}
\thanks{Xiaolan Liu is with School of Electrical, Electronic and Mechanical Engineering, University of Bristol, U.K. (Email:xiaolan.liu@bristol.ac.uk).}
\thanks{Sangarapillai Lambotharan is with Institute for Digital Technologies, Loughborough University, U.K. (Email:S.Lambotharan@lboro.ac.uk).}
}
%\thanks{Manuscript created October, 2020; This work was developed by the IEEE Publication Technology Department. This work is distributed under the \LaTeX \ Project Public License (LPPL) ( http://www.latex-project.org/ ) version 1.3. A copy of the LPPL, version 1.3, is included in the base \LaTeX \ documentation of all distributions of \LaTeX \ released 2003/12/01 or later. The opinions expressed here are entirely that of the author. No warranty is expressed or implied. User assumes all risk.}
}

%\markboth{Journal of \LaTeX\ Class Files,~Vol.~18, No.~9, September~2020}%
%{How to Use the IEEEtran \LaTeX \ Templates}

\maketitle

\begin{abstract}
In 6G communications, it is envisioned to equip the traditional access point (AP) with sensing capability to fully benefit the existing wireless communication infrastructures. Thus, sensing-assisted communication has attracted significant attention from both industry and academia. However, most existing works focused on sensing-assisted communication in line-of-sight (LoS) scenarios due to sensing limitations, where the sensing target (ST) and communication user equipment (UE) remain the same. In this paper, we propose a general sensing-assisted channel estimation framework in the distributed multiple-input and multiple-output (DMIMO) network and consider a scenario where the ST and UE are different entities. In addition, ST is a moving target (e.g. a robot)  which causes channels between APs and UEs to vary due to changes in the reflection paths of the indoor environment. Therefore, we let multiple APs to jointly sense the position of the ST, which will be incorporated in a Ray tracing model to obtain a more accurate estimate of the channels from APs to UEs for both the LoS and non-line-of-sight (NLoS)  scenarios. Simulation results demonstrate that our proposed sensing-assisted communication framework achieves a much higher channel estimation accuracy and downlink throughput compared to the traditional least-square (LS) channel estimation. More importantly, the feasibility of the proposed framework has been validated to guarantee the stringent channel estimation accuracy requirement in the DMIMO network.

%especially considering the promising distributed multiple-input multiple-output (MIMO) networks. This paper focuses on formulating a throughput maximization problem to improve channel estimation accuracy for users (UE) by sensing the information of non-cooperative moving targets (ST) (i.e., the object is not in communication) in the communication environment, and then raytracing method is used to calculate and identify the ST-associated non-line-of-sight (NLoS) propagation paths that can enhance the channel coefficient. 
\end{abstract}


\section{Introduction}
The Joint sensing and communication (JSC) system has been recognized as a key technology to support various emerging scenarios in 6G networks, e.g., autonomous driving, digital twin,  and extended reality.  It enables sensing and communication to efficiently share the hardware and wireless resources in the existing communication infrastructure \cite{liu2022survey, demirhan2023cell}. Shifted from transmission waveform design, multiple input and multiple output (MIMO) beamforming optimization, and joint resource allocation optimization in single JSC access point (AP) scenario, the more advanced distributed MIMO (DMIMO) scenario has received significant attention recently, where multiple JSC APs are operating in the same frequency band, and time to achieve higher sensing accuracy and communication performance \cite{demirhan2023cell,liu2023cell}. Prior work on JSC can be mainly categorized into: 1) \textit{sensing-assisted communication} including sensing-assisted beam tracking/training \cite{Mu2021,Cazzella2024}, resource allocation and channel estimation \cite{Huang2021}; 2) \textit{communication-assisted sensing} consisting of distributed/networked sensing\cite{Liu2022}, and channel state information (CSI)-enhanced sensing \cite{Lu2024}; and 3) \textit{joint sensing and communication} including waveform design \cite{Liu2022}, and beamforming optimization \cite{demirhan2023cell}.

In existing works on sensing-assisted communication, the authors mainly focused on the line-of-sight (LoS) communication scenario with a single JSC AP, where the sensing target (ST) and the communication user equipment (UE) are assumed to be the same object. In \cite{Huang2021}, a MIMO radar is first deployed to measure the azimuth information of moving vehicles, which is used to develop channel estimation. Although a reflection model for millimeter wave propagation is presented for environment perception, and the non-line-of-sight (NLoS) mmWave channel can be estimated based on the information of the located reflectors \cite{jiao2018indoor}. The authors merely focused on the static sensing target with a single JSC AP instead of the DMIMO network. More importantly, the sensing-assisted channel estimation in a dynamic wireless environment remains unsolved.

Motivated by the above challenge, we consider a DMIMO network where the moving ST causes varying channel conditions between the APs and UEs. Due to the stringent requirement of channel estimation accuracy in the DMIMO network, we focus on enhancing the channel estimation between APs and UEs by utilizing the sensing information of the ST. Specifically, we assume that there is a dedicated sensing slot for APs to jointly perform sensing capability in each transmission frame of future 6G communications \cite{zhang2022}. Then, the proposed sensing-assisted channel estimation framework constructs the estimated channel in a dynamic wireless environment by using the Ray-tracing method to calculate both the LoS and NLoS propagation paths. The main contributions are as follows
\begin{itemize}
\item Different from existing works that mainly focus on LoS scenarios with a single JSC AP, we proposed a general sensing-assisted channel estimation framework for the DMIMO network, which captures the characteristics of both the LoS path and NLoS path via the Ray-tracing technique;
\item The proposed sensing-assisted channel estimation framework is able to function well in a dynamic environment instead of a static environment, which achieves a higher channel correlation coefficient and downlink throughput under various transmission power and number of APs compared to the traditional least-square (LS) channel estimation algorithm.
\end{itemize}
\begin{comment}
        \item Compared to the traditional channel estimation method, the proposed sensing-assisted communication framework enhances the communication performance under NLoS scenario; (Fig.1: throughput evolution along with time, enlarge the NLoS part due to the sensing target,should we allocate same resource for sensing and communication)
    \item The proposed sensing-assisted communication achieves higher performance gain with a higher number of participated base stations under high mobility scenario. (Fig.2: different number of base stations and sensing target speed)
    \item We evaluate the trade-off between the sensing resource and communication resource allocation, which sheds light on the resource allocation optimization for the future JSC system. (Fig.3: performance with different number of sensing slots)
\end{comment}

The rest of this letter is organized as follows. Section II presents the system model and problem formulation. Section III describes the sensing-assisted channel estimation framework.
Section IV presents simulation results followed by the conclusions in Section V.
%1. novel transmission frame structure
%2. Sensing target and communication users are different objects 
%3. conventional LoS scenario, only care about the amulizth,  NLoS scenario, we need to include %all of the propagation paths, sensing environment and calculate propagation paths. 
%dynamic environment with moving sensing target. 

% All the MIMO schemes are heavily dependent on perfect channel state information (CSI), thus requiring ultra-high channel estimation accuracy \cite{jiao2018indoor}.  especially in massive antennas and multiple UEs cases.

\section{System Model and problem formulation}

\begin{figure}[h!]
    \centering
    \includegraphics[scale=0.8]{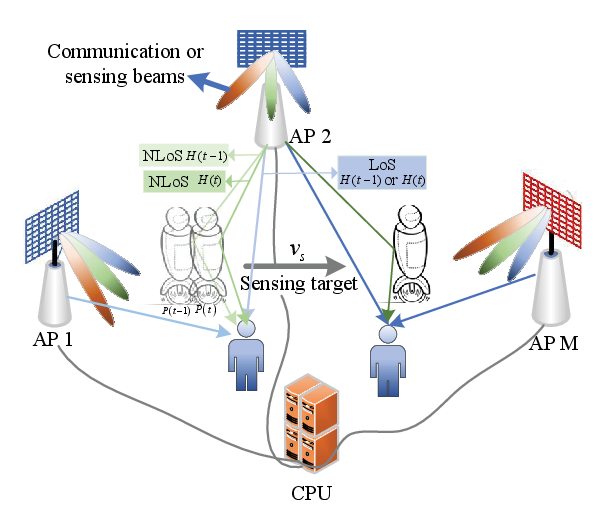}
    \caption{System model}
    \label{system_model}
\end{figure}

We consider a DMIMO JSC system with $M$ APs, where each AP $m\in\mathcal{M}=\{1,...,M\}$ is equipped with $N_t$ transmitting antennas and $N_r$ receiving antennas in full-duplex mode. All the APs are connected to a central processing unit (CPU) that can jointly perform data processing. 
We assume there are $U$ UEs, where each UE $u\in\mathcal{U}=\{1,...,U\}$ has $N_u$ antennas, and one mobile ST $s$ moving with a speed $v_s$ towards a certain direction. It is noted that the ST is modeled as an extended target, whose length $l_s$ and width $w_s$ are assumed known. As shown in Fig.~\ref{system_model}, we consider each transmission frame $\mathcal{F}_j$ consists of $N$ slots, i.e., $\mathcal{F}_j =\{0,...,N-1\} $, where the length of each slot is $\Delta T$. In the sensing slot of each frame, i.e., $\mathcal{F}_{j,0}$, $M$ APs operate in the monostatic radar mode, where each AP $m$ transmits the sensing signal and receives the reflections/scattering from the ST based on full-duplex mode. In the communication slots of each frame, i.e., $\mathcal{F}_{j,\{1,...,N-1\}}$, all $M$ APs serve the $U$ UEs simultaneously \cite{zhang2022}.

\subsection{Sensing Signal Model}

We consider that each AP is equipped with a uniform linear array (ULA) \cite{Du2023integrated}, and transmits the signal towards the center of the ST using knowledge of the predicted position of the ST from the previous iteration in a tracking setup. We assume the $K$ resolvable scatterers are uniformly distributed in the ST geometry. Hence, once the coordinates of these scatterers are localized,
the coordinates of the centroid can be uniquely determined. In the $\mathcal{F}_{j,0}$ sensing slot, the received sensing signal at the $m^{\mathrm{th}}$ AP can be represented as
\begin{equation}
\begin{aligned}
\textbf{r}_{m}^{j,0} (t) = \kappa \sqrt{p_{m}}\sum ^{K}_{k=1} \beta_{m,k}^{j,0} e^{j2\pi \mu_{m,k}^{j,0} t}...\\
...\textbf{b}(\theta_{m,k}^{j,0})\textbf{a}^{\dagger}(\theta_{m,k}^{j,0})\textbf{f}^{j,0} {s}(t-\tau_{m,k}^{j,0}) + \textbf{z}(t),
\end{aligned}
\end{equation}
where $\kappa = \sqrt{N_t N_r}$ is the array gain, $p_{m}$ is the transmission power, $\beta_{m,k}^{j,0}$, $\mu_{m,k}^{j,0}$ and  $\tau_{m,k}^{j,0}$ represent the complex reflection coefficient, Doppler frequency, and round-trip delay of $k^{th}$ scatter. The $\mathbf{b}(\theta_{\mathrm{AOA}})$ and  $\mathbf{a}(\theta_{\mathrm{AOD}})$ represent the array response vector corresponding to the reception and transmission of signals at $\theta_{\mathrm{AOA}}$ and $\theta_{\mathrm{AOD}}$ as shown in \eqref{eq:AOA} and \eqref{eq:AOD}, where $\theta_{\mathrm{AOD}} = \theta_{\mathrm{AOA}} = \theta_{m,k}^{j,0} $ due to monostatic radar setting. The $s(t)$ is the transmitted sensing signal, and $\textbf{z}(t)$ is the additive white Gaussian noise with zero mean and variance of $\sigma^2$, i.e., $\textbf{z}(t) \sim \mathcal{C}\mathcal{N}(\textbf{0}_{N_r}, \sigma^2\mathbf{I}_{N_r \times 1})$.

\begin{equation}
\textbf{b}(\theta_\mathrm{AOA}) = \frac{1}{\sqrt{N_r}} [1, e^{-j\pi\cos(\theta_\mathrm{AOA})},...,e^{-j\pi(N_r - 1)\cos(\theta_\mathrm{AOA})}]^T.
\label{eq:AOA}
\end{equation}

\begin{equation}
\textbf{a}(\theta_\mathrm{AOD}) = \frac{1}{\sqrt{N_t}} [1, e^{-j\pi\cos(\theta_\mathrm{AOD})},...,e^{-j\pi(N_t - 1)\cos(\theta_\mathrm{AOD})}]^T.
\label{eq:AOD}
\end{equation}

The complex reflection coefficient $\beta_{m,k}^{j,0}$ is determined by the signal
propagation distance and the radar cross section (RCS) of the
ST, expressed as:
\begin{equation}
    \beta_{m,k}^{j,0} = \frac{\epsilon^{j,0}_{m,k}}{(2d^{j,0}_{m,k})^2},
\end{equation}
where $d^{j,0}_{m,k}$ is the distance between the scatter and AP, and the RCS $\epsilon^{j,0}_{m,k}$ follows zero mean and unit variance Gaussian distribution.

The $\textbf{f}^{j,0}$ represents the beamforming vector, which is written as:
\begin{equation}
\textbf{f}^{j,0} = \textbf{a}\left(\hat{\phi}_{(j,0)|(j-1,0)}^{m}\right),
\end{equation}
where $\hat{\phi}_{(j,0)|(j-1,0)}^{m}$ represents the predicted angle of the center of ST based on estimated position and velocity of ST in the $\mathcal{F}_{j-1.0}$ sensing slot.

\subsection{Sensing Measurement Model}
We consider the classic radar processing technique, i.e., matched-filtering, to estimate the round-trip delay and Doppler frequency of the ST. This is implemented by correlating the received signal with the transmitted waveforms at different delays and performing an FFT, as in all radar signal processing, to generate Delay-Doppler maps. In this way, the moving target would be clearly distinguishable from clutter, which would normally appear closer to zero Doppler in the Delay-Doppler domain. The processed signal can be represented as:
\begin{equation}
   \begin{aligned} &\widetilde {\mathbf {r}}_{m}^{j,0}=\kappa \sqrt {p_{m}} \sum ^{K}_{k=1} \beta _{m,k}^{j,0} \mathbf {b} \left ({\theta _{m,k}^{j,0}}\right) \mathbf {a}^{\dagger} \left ({\theta _{m,k}^{j,0}}\right) \mathbf {a} \left(\hat{\phi}_{(j,0)|(j-1,0)}^{m}\right) \\&\times \int ^{\Delta T}_{0} s \left ({t-\tau _{m,k}^{j,0}}\right) s^{*} \left ({t-\tau }\right) e^{-j2\pi (\mu -\mu _{m,k}^{j,0}) t} dt +\widetilde {\mathbf {z}}_{r} \\=&\kappa  \sqrt {p_{m}} \sqrt {G} \sum ^{K}_{k=1} \beta _{m,k}^{j,0} \mathbf {b} \left ({\theta _{m,k}^{j,0}}\right) \mathbf {a}^{\dagger} \left ({\theta _{m,k}^{j,0}}\right) \mathbf {a}\left(\hat{\phi}_{(j,0)|(j-1,0)}^{m}\right) \\&\times \,\bar {\delta } \left ({\tau -\tau _{m,k}^{j,0}; \mu -\mu _{m,k}^{j,0} }\right) +\widetilde {\mathbf {z}},\end{aligned} 
\end{equation}
where $\Delta{T}$ is the length of the sensing slot, and $G$ represents the matched-filtering gain. The $\bar {\delta } \left ({\tau; \mu }\right)$ represents the normalized matched-filtering output function, which has a narrow mainlobe property in delay domain and Doppler domain to ensure high resolution, i.e., $\bar {\delta } \left ({\tau; \mu }\right) = 1$ when $\tau = 0$, and $\mu = 0$.

Therefore, the measured round-trip delay and Doppler frequency can be represented as:
\begin{equation}
    \begin{aligned} \widehat {\tau }_{m,k}^{j,0} = \frac {2d_{m,k}^{j,0}}{c} + z_{\tau _{m,k}^{j,0}}, \quad \widehat {\mu }_{m,k}^{j,0} = \frac {2v_{s}\cos (\theta _{m,k}^{j,0})f_{c}}{c} + z_{\mu _{m,k}^{j,0}}. \end{aligned}
\end{equation}

%Based on the known RCS and distance:
%\begin{equation*} \beta _{k,n} = \frac {\varepsilon _{k,n}}{(2d_{k,n})^{2}} = \frac {\varepsilon _{k,n}}{c^{2}\tau ^{2}_{k,n}},\tag{9}\end{equation*}

%It can also be calculated based on Eq 12 in \cite{Jiao2019}

The $\theta _{m,k}^{j,0}$ can be estimated by the multiple
signal classification (MUSIC) algorithm, whose measurements are expressed as
\begin{equation} \widehat \theta _{m,k}^{j,0} = \theta _{m,k}^{j,0} + z_{\theta _{m,k}^{j,0}},\end{equation}
where the $z_{\tau _{m,k}^{j,0}}$, $z_{\mu _{m,k}^{j,0}}$ and $z_{\theta _{m,k}^{j,0}}$ are additive noises with zero means and variances of $\sigma^2_{\tau,j,m,k}$, $\sigma^2_{\mu,j,m,k}$, and $\sigma^2_{\theta,j,m,k}$.

Then the $\sigma^2_{\tau,j,m,k}$, $\sigma^2_{\mu,j,m,k}$, and $\sigma^2_{\theta,j,m,k}$ can be represented as:
\begin{equation} 
\sigma ^{2}_{i,j,m,k} = \frac {a^{2}_{i}\sigma ^{2}}{p_{m} G|\kappa  \beta _{m,k}^{j,0}|^{2}\left \vert{ \varrho_{m,k}^{j,0} }\right \vert ^{2}}, i = {\tau,\mu,\theta},
\end{equation}
where $\varrho_{m,k}^{j,0} = \mathbf {a}^{\dagger} \left ({\theta _{m,k}^{j,0}}\right) \mathbf {a} \left(\hat{\phi}_{(j,0)|(j-1,0)}^{m}\right)$ represents the beamforming gain, and $a_{i}$ are constants related to the system configuration, signal designs, and algorithms.

Finally, we can obtain the measurements of the coordinates of centroid, the angle of the centroid, and the velocity of the ST, which can be represented as
\begin{equation}
\begin{aligned} \widehat {x}_{j,0}=&\frac {1}{MK}\sum ^{M}_{m=1}\sum ^{K}_{k=1}{c\widehat {\tau }_{m,k}^{j,0}\cos \widehat {\theta }_{m,k}^{j,0}}/2,\\ \widehat {y}_{j,0}=&\frac {1}{MK}\sum ^{M}_{m=1}\sum ^{K}_{k=1}{c\widehat {\tau }_{m,k}^{j,0}\sin \widehat {\theta }_{m,k}^{j,0}}/2,\\
\widehat {v}_{j,0} =& \frac {c}{2f_{c}} \frac{1}{M}\sum ^{M}_{m=1}\frac {\sum ^{K}_{k=1}\widehat {\mu }_{m,k}^{j,0}\cos (\widehat {\theta }_{m,k}^{j,0})/\sigma ^{2}_{\mu,j,m,k} }{\sum ^{K}_{k=1}\cos ^{2}(\widehat {\theta }_{m,k}^{j,0})/\sigma ^{2}_{\mu,j,m,k} }.\end{aligned} 
\label{eq:estimated_position}
\end{equation}

\subsection{Communication Model}
In the DMIMO setting, we denote the estimated communication channel between UE $u$ and AP $m$ in $\mathcal{F}_{j,n},  n \in \{1,...,N-1\}$ communication slot as $\mathbf{\widehat{h}_{u,m}^{n}} \in \mathbb{C}^{N_u \times N_t}$. We assume a central server collects all the estimated channel $\mathbf{\widehat{h}_{u,m}^{j,n}} \in \mathbb{C}^{N_u \times N_t}$ and stacks the channels between UE $u$ and all the APs to obtain $\mathbf{\widehat{h}_{u}^{j,n}} \in \mathbb{C}^{N_u \times MN_t}$. Without loss of generality, we consider the well-known zero-forcing (ZF) beamforming algorithm, where the complete estimated channel information is represented as $\mathbf{\widehat{H}}^{j,n} = [\mathbf{\widehat{h}}_1^{j,n}; ...;\mathbf{\widehat{h}}_U^{j,n}]
\in \mathbb{C}^{UN_u \times MN_t}
$. Therefore, the ZF beamforming weights $W^{j,n}$ are calculated as 
\begin{equation}
    \mathbf{W}^{j,n} = {\mathbf{\widehat{H}}^{{j,n}\dagger}}(\mathbf{\widehat{H}}^{j,n}\mathbf{\widehat{H}}^{{j,n}\dagger})^{-1},
\end{equation}
where $\mathbf{W}^{j,n}=\left[\mathbf{w}^{j,n}_1, ..., \mathbf{w}^{j,n}_U\right] \in \mathbb{C}^{MN_t \times UN_u}$, and $\mathbf{w}^{j,n}_u = \left[\mathbf{w}_{u,1};...;\mathbf{w}_{u,M}\right] \in \mathbb{C}^{MN_t \times N_u}$ represents the beamforming weights of each UE.

Then, considering a block fading channel model, we can present the received signal at UE $u$ in each communication slot $\mathcal{F}_{j,n}$ as
\begin{equation} 
\begin{aligned}
\mathbf{y_{u}^{j,n}} &= \underbrace{\sqrt{p_u}\mathbf{h_{u}^{j,n}} \frac{\mathbf{w_{u}^{j,n}}}{||\mathbf{w_{u}^{j,n}}||_{\mathrm{F}}} \mathbf{x_u}}_{\mathrm{Desired Signal}}  + \\&\underbrace{\sum_{u^{'} \in \mathcal{U} \setminus \{u\}}\sqrt{p_{u^{'}}}\mathbf{h_{u}^{j,n}} \frac{\mathbf{w_{u^{'}}^{j,n}}}{||\mathbf{w_{u^{'}}^{j,n}}||_\mathrm{F}}\mathbf{x_{u^{'}}}}_{\mathrm{Interference}} +\mathbf{z_u},
\end{aligned}
\end{equation}
where $p_u = M * p_{m}$ represents the allocated transmission power for each user in DMIMO setting, $\mathbf{x}_u \in \mathbb{C} ^{N_u \times 1} $ represents the transmitted user data, and $\mathbf{z_u}$ is the Gaussian noise with zero mean and $\sigma^2$ variance.

\begin{comment}
We consider the well-known maximum ratio transmission (MRT), hence, the beamforming weights are calculated as
\begin{equation} 
\begin{aligned}
w_{u, m}^{n} = {\widehat{h}_{u,m}^{n}}^{H},
\end{aligned}
\end{equation}
where $\widehat{h}_{u,m}^{n}$ is the estimated channel between user $u$ and AP $m$.

Therefore, the SINR can be obtained as follows:
\begin{equation}\label{eq:SINR}
\begin{aligned}
\mathrm{SINR}_{u}^{n} &= \frac{||h_{u}^{n} w_{u}^{n}||^{2}_{F}}{\sum_{u^{'} \in \mathcal{U} \setminus \{u\}}||h_{u}^{n} w_{u^{'}^{n}}||^{2}_{F}+\sigma^2}
\\& = \frac{||h_{u}^{n} {\widehat{h}_{u}^{n}}^{H}||^{2}_{F}}{\sum_{u^{'} \in \mathcal{U} \setminus \{u\}}||h_{u}^{n} {\widehat{h}_{u^{'}}^{n}}^{H}||^{2}_{F}+\sigma^2}.
\end{aligned}
\end{equation}    
\end{comment}

\subsection{Problem Formulation }
To enhance downlink transmission throughput in our considered DMIMO scenario, we aim to maximize the downlink throughput by optimizing the estimated channel state information, which can be formulated as follows:
\begin{equation}\label{eq:P1}
\begin{aligned}
    (\text {OP1})\!:\quad \mathop {\textbf {max}}\limits_{\mathbf{\widehat{h}}_u^{j,n}}\sum_{j=1}^{\infty}\sum_{n =1}^{N}\sum_{u \in \mathcal{U}}&\log2\det(\mathbf{R}_\mathrm{S}^{u,{j,n}} + \mathbf{R}_\mathrm{I}^{u,{j,n}}) \\&- \log2\det(\mathbf{R}_\mathrm{I}^{u,{j,n}}),
\end{aligned}
\end{equation}
where 
\begin{equation}
    \mathbf{R}_\mathrm{S}^{u,{j,n}} = p_u \mathbf{h_{u}^{j,n}} \frac{\mathbf{w_{u}^{j,n}}\mathbf{w_{u}^{{j,n}\dagger}}}{||\mathbf{w_{u}^{j,n}}||_{\mathrm{F}}^2}\mathbf{h_{u}^{{j,n}\dagger}},
\end{equation}
and
\begin{equation}
    \mathbf{R}_\mathrm{I}^{u,{j,n}} = \sum_{u^{'} \in \mathcal{U} \setminus \{u\}}{p_{u^{'}}}\mathbf{h_{u}^{j,n}} \frac{\mathbf{w_{u^{'}}^{j,n}}\mathbf{w_{u^{'}}^{{j,n}\dagger}}}{||\mathbf{w_{u^{'}}^{j,n}}||_\mathrm{F}^2}\mathbf{h_{u}^{{j,n}\dagger}} + \sigma^2\mathbf{I}.
\end{equation}

%\subsection{Problem Formulation for Throughput Maximization}
%Therefore, the throughput maximization problem 

%to maximize the communication throughput, the problem can be formulated as follows
% 	\begin{equation}\label{eq:P1} (\text {OP1})\!:\quad \mathop {\textbf {max}}\sum_{u \in \mathcal{U}}\log (1 + \mathrm{SINR}_{u}^{n}),
% 	\end{equation}
% where $\mathrm{SINR}_{u}^{n}$ is provided in Eq.~(\ref{eq:SINR}).

% To simplify the coupled interference between UEs, we assume that low channel correlation between UEs, which means $||h_{u} \widehat{h}_{u^{'}}^{H}||^{2}_{F} \approx 0$. Therefore, the optimization problem in Eq.~(\ref{eq:P1}) can be transformed into Eq.~(\ref{eq:P2})
% 	\begin{equation}\label{eq:P2} (\text {OP2})\!:\quad \mathop {\textbf {max}}\sum_{u \in \mathcal{U}}||h_{u}^{n} \widehat{h}_{u}^{n}^{H}||^{2}_{F},
% 	\end{equation}
% \textbf{which aims to improve the channel estimation accuracy.}

\section{Sensing-Assisted Channel Estimation Framework}

\begin{figure}[t]
    \centering
    \includegraphics[scale=1]{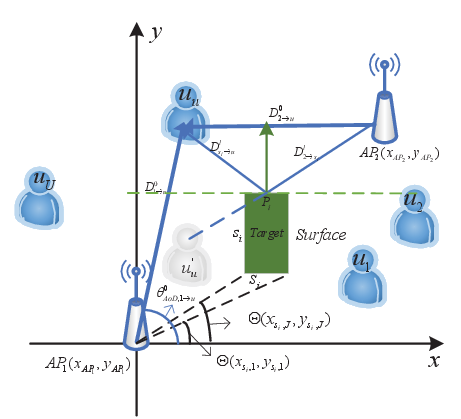}
    \caption{Propagation reflection model between APs and UEs }
    \label{reflection model}
\end{figure}

In this section, the proposed sensing-assisted channel estimation scheme will be presented. First, we consider that one sensing slot is allocated in each transmission frame to detect the position and velocity of the ST.  Then, we utilize the estimated position of the ST to calculate the propagation paths between each AP and UE, where both LoS and NLoS propagation paths are considered via Ray-tracing method. %Hence, new channel information can be updated in each transmission frame, which then contributes to alleviating the channel aging effect caused by the dynamic surrounding environment. 
 %Based on the reciprocity, the AP allocates the periodic uplink pilot resource to perform channel estimation, where the period of pilot transmission is assumed to be $N$ frames due to limited pilot resources.
 
 %Therefore, to alleviate the channel aging effect caused by the dynamic surrounding environment, 
%we propose to update the channel estimation in each frame based on the obtained sensing information in the sensing slot.
% \begin{itemize}
%     \item Obtain sensing information of ST with the allocated sensing slot
%     \item Calculate both LoS and NLoS propagation paths using ray tracing method to obtain the propagation gain
%     \item 
% \end{itemize}

\begin{comment}
Then, the corresponding unit norm array response vectors are defined as 

\begin{equation}
\textbf{a}_{m,l}(\theta_{m,l}) = \frac{1}{\sqrt{N_m}} [1, e^{-j\pi\cos(\theta_{m,l})},...,e^{-j\pi(N_m - 1)\cos(\theta_{m,l})}]^T, \in \mathbb{C}^{N_m}
\end{equation}

\begin{equation}
\textbf{b}_{u,l}(\theta_{u,l}) = \frac{1}{\sqrt{N_r}} [1, e^{-j\pi\cos(\theta_{u,l})},...,e^{-j\pi(N_r - 1)\cos(\theta_{u,l})}], \in \mathbb{C}^{N_u}.
\end{equation}
\end{comment}
\subsection{Multipaths Channel Model} 

We consider the channel between any AP $m$ and the UE $u$ comprises $L \geq 1$ paths, where the first one is the LoS path and the other $L-1$ paths are associated with the reflection of ST. Here, $L$ satisfies $L \ll min(N_u, N_t)$ in mmWave propagation, and only single-bounce reflections are considered. We assume the APs only perform channel estimation in each sensing slot, i.e., $\mathcal{F}_{j,0}$, where the estimated channel remains the same between two sensing slots, i.e., $\mathbf{\widehat{h}_{u,m}^{j,0}}=\mathbf{\widehat{h}_{u,m}^{j,1}}=...=\mathbf{\widehat{h}_{u,m}^{j,N-1}}$. 

Let $\widehat{\theta}^{j,0}_{m,u,l}$ and $\widehat{\theta}^{{j,0}}_{u,m,l}$, $0 \leq l \leq L$, be the estimated AOD and AOA of $l^{th}$ communication path based on the estimated ST position, where $l=0$ represents the LoS path. Therefore, the channel coefficient between any AP $m$ and the user $u$ can be expressed as:
\begin{equation}
\mathbf{\widehat{h}_{u,m}^{j,0}}= \sum_{l=0}^{L} I_{m \to u}^l \sqrt{N_t N_u}  \widehat{\alpha}_{m \to u} ^l \textbf{c}(\widehat{\theta}^{{j,0}}_{u,m,l}) {\textbf{a}(\widehat{\theta}^{j,0}_{m,u,l})}^\dagger,
\end{equation}
where 
\begin{equation}
    \textbf{c}(\theta_\mathrm{AOA}) = \frac{1}{\sqrt{N_u}} [1, e^{-j\pi\cos(\theta_\mathrm{AOA})},...,e^{-j\pi(N_u - 1)\cos(\theta_\mathrm{AOA})}]^T.
\end{equation}

It is noted that $I_{m \to u}^l  = 1$ indicates the propagation path $l$ exists, otherwise not. In the following sections, we will present how to calculate the propagation gain $\widehat{\alpha}_{m \to u} ^l$ under the condition of the LoS path and NLoS path. 
%= \sum\limits_{l = 0}^L {{h_{l}(t)}\delta(t-\tau_l)}
 % \begin{equation}
 % {h_l(t)} \buildrel \Delta \over = \sqrt {{N_u}{N_m}} \beta_l {\alpha_l}{e^{j2\pi f_D cos(\theta_{AoA, m \to R})}}{\textbf{a}_{u,l}} {\textbf{b}_{u,l}^H}. 
 % \end{equation}
%where $\tau_l = \frac{D_l}{c}, D_l=D_{m \to re} +D_{re \to u}$ is the delay of the NLoS path $l^{th}$. 

% $\alpha_l$, $\beta_l$ is the reflection propagation gain and the complex gain of the $l^{th}$ path.  Considering the Doppler effect on the NLoS path caused by the mobile ST, the channel matrix between the AP and ST will include an extra term $e^{i2\pi f_D cos(\theta_{AoA,m \to re})}$ \cite{caslchannel}. 

\subsection{LoS Path}
It is noted that the mobility of the ST causes variations of communication links between UEs and APs, and may lead to blockage of the LoS path. Therefore, the existence of LoS path is determined based on the estimated position of ST in \eqref{eq:estimated_position} and prior width $w_s$ and length $l_s$ information.

%we have to detect if there are any LoS and NLoS propagation paths. The ST is assumed to be rectangular, if we know the position of the ST, represented with the centroid as $(x_s,y_s)$, width $W_s$ and length $L_s$, the four surfaces (i.e., reflective surface) of the ST can be obtained. The four surfaces are represented as $\{s_i, i=1,2,3,4\}$. 
% \subsection{mmWave Channel Estimation Model}
% After sensing the target, the mmWave APs possess the information of the moving target. Once the users' positions are known to the mmWave APs, they can calculate all the propagation paths from them to the users. Let the user $u$ have the coordinate $(x_u,y_u)$. Due to the mobile target, the communication links between the APs and the users may be blocked at some time. Therefore,
% to estimate channel state information, we have to consider both line-of-sight (LoS) and non-LoS (NLoS) propagation paths. 

The LoS paths between APs and UE exist only if the ST is not blocking the communication link, i.e., they are not blocked by any surface of the ST, like the LoS link between AP ${AP}_1$ and UE $u_u$ shown in Fig. \ref{reflection model}, where four surfaces of the ST are represented as $\{s_i, i=1,2,3,4\}$. As shown in Fig. \ref{reflection model}, each surface can induce an angular range defined as 
\begin{equation}
\Omega_{{s_i}}  \buildrel  \over = \left\{ {\begin{array}{*{20}{c}}
{[\Theta(x_{{s_i},1}, y_{{s_i},1}), \Theta(x_{{s_i},J}, y_{{s_i},J})),}\\
{\Theta(x_{{s_i},1}, y_{{s_i},1}) < \Theta(x_{{s_i},J}, y_{{s_i},J})}\\
{[\Theta(x_{{s_i},1}, y_{{s_i},1}), 2\pi) \cup [0, \Theta(x_{{s_i},J}, y_{{s_i},J})),}\\
{\Theta(x_{{s_i},1}, y_{{s_i},1}) > \Theta(x_{{s_i},J}, y_{{s_i},J}).}
\end{array}} \right.
\end{equation}
where $\Theta(x_{{s_i},1}, y_{{s_i},1}),\Theta(x_{{s_i},J}, y_{{s_i},J}) $ are the angles of the two endpoints of the surface $s_i$. The calculation of $\Theta(x,y)$ can be found in \cite{Jiao2019}.

% To judge the existence of the LoS path, we define 
% \begin{equation}
% \Theta(x,y)  \buildrel \Delta \over = \left\{ {\begin{array}{*{20}{c}}
% {\text{mod}(\arctan (\frac{y-y_{m}}{x-x_{m}}),2\pi),\; x \ge x_{m}}\\
% {\arctan (\frac{y-y_{m}}{x-x_{m}}) + \pi, \;\;\;\;\;\;\;\;x < x_{m}}\\
% {\pi- \frac{y-y_{m}}{2\lvert y-y_{m}\rvert} \pi, \;\;\;\;\;\;\;\;\;\;\;\;\;\;\;\;x = x_{m}}
% \end{array}} \right.
% \end{equation}

%and thus we have to calculate the NLoS paths to increase the propagation gain \cite{jiao2018indoor}. 
Therefore, the LoS paths exist, i.e., $I_{m \to u}^0  = 1$, if and only if $\Theta(x_u,y_u)$ does not fall within any range $\Omega_{s_i}, i=\{1,2,3,4\}$. 
Then, the corresponding propagation gain of the LoS path $ \widehat{\alpha} _{m \to u}^{0}$ is calculated as:
\begin{equation}
\begin{array}{l}
\widehat{\alpha} _{m \to u}^{0} = e^{j\frac{2\pi D_{m \to u}^{0}}{\lambda}}\sqrt {\min (\frac{A_u^0}{{W(\widehat{\theta} _{m,u,0},{N_t})D_{m \to u}^{0}}},1)}, \\
D_{m \to u}^{0} = \sqrt {{{({\widehat{x}_u} - {x_m})}^2} + {{({\widehat{y}_u} - {y_m})}^2}}
\end{array}
\end{equation}
where $A_u^0$ is the effective user aperture $A_u^0 = \lambda + (N_{u} -1)\lambda \lvert sin(\widehat{\theta} _{u,m,0})\rvert$,  and
$D_{m \to u}^{0}$ is the estimated distance between the AP $m$ and the user $u$. 

%$\widehat{\theta} _{AoA,m \to u}^{0}, \widehat{\theta} _{AoD,m \to u}^{0}$ are the AOA and AOD of this path between the AP $m$ and UE $u$. 

% Therefore, the channel matrix between the AP and the user with the LoS path is given by
% \begin{equation}
% {h^{0}(t)} \buildrel \Delta \over = \sqrt {{N_u}{N_m}} \beta_0 {\alpha _{m \to u}^{0}}{\textbf{a}_{u}^0} {{{\textbf{b}_{m}^0}^H}}. 
% \end{equation}

% As defined above, if there is a LoS path, denoted as $l=0$, existing between the AP $m$ and the user $u$, the angles of departure and arrival of this path can be obtained by \cite{jiao2018indoor}
% \begin{equation}
% \begin{array}{l}
% \theta _{AoD,m \to u}^{0} = \Theta ({x_{m \to u}},{y_{m \to u}})\\
% \theta _{AoA,m \to u}^{0} = \text{mod} (\Theta ({x_{m \to u}},{y_{m \to u}}) - {\varphi _{ u}} + \pi, 2\pi) 
% \end{array}  
% \end{equation}
% where $\varphi _{u}$ is the array orientation of user $u$. 

%From \cite{jiao2018indoor}, 
\subsection{NLoS Path}
Considering the reflective (i.e., NLoS) paths caused by ST, the NLoS path exists, i.e., $I_{m \to u}^l  = 1$, if and only if the AP and the mirror UE $u_u^{\prime}$ over the surface $s_i$ have an interaction point, and the reflective path is not blocked by any other surfaces. For instance, in Fig. \ref{reflection model}, one NLoS link exists between ${AP}_2$ and UE $u_u$ with the reflection point $P_i$ on surface $s_i$. Therefore, the reflection coefficient (propagation gain) $\widehat{\alpha} _{m \to u}^{l}$ can be calculated as \eqref{alpha_NLoS}. 

\begin{figure*}
\begin{equation}\label{alpha_NLoS}
\begin{array}{l}
\widehat{\alpha} _{m \to u}^{l} = {e^{ - j({{\hat \Phi }_{s_i}^l}- \frac{2\pi (D_{m \to {s_i}}^{l} + D_{{s_i} \to u}^{l})}{\lambda})}} \\
\times \sqrt {\eta \left( {\min \left( {\frac{A_u^l}{{W(\widehat{\theta} _{m,u,l},{N_t})(D_{m \to {s_i}}^{l} + D_{{s_i} \to u}^{l})}},1} \right){{\hat R}_{s}^l} + \frac{{{{\sin }^2}\left( {\Theta (\frac{1}{{{a_l}}},1) - \Theta (x_{m \to u}^{l},y_{m \to u}^{l})} \right){A_u^l}}}{{\sqrt {4{{\left( {D_{{s_i} \to u}^{l}} \right)}^2} + {{A_u^l}^2}} }}{{\hat R}_{d}^l}} \right)}, \\
D_{m \to {s_i}}^{l} = \sqrt {(x_{s_i}^l-x_m)^2 + (y_{s_i}^l-y_m)^2} ,{\kern 1pt} {\kern 1pt} {\kern 1pt} {\kern 1pt} {\kern 1pt} D_{{s_i} \to u}^{l} = \sqrt {(x_{s_i}^l-x_u)^2 + (y_{s_i}^l-y_u)^2}.
\end{array}  
\end{equation}
\end{figure*}

In (\ref{alpha_NLoS}), $D_{m \to {s_i}}^{l}$ and $D_{{s_i} \to u}^{l}$ are the distances from the reflection surface $s_i$ to the AP $m$ and the user $u$, and $\{{\hat \Phi }_{s_i}^l, {\hat R}_{s}^l, {\hat R}_{d}^l\}$ denotes the reflection property of the reflective surface. ${\hat \Phi }_{s_i}^l$ is the phase shift determined by the reflection property, ${\hat R}_{s}^l \in (0,1)$ and ${\hat R}_{d}^l \in (0,1)$ are the sepecular reflectance and diffuse reflectance of the reflective surface, respectively.

\begin{table}[h!]
	\centering  
	\caption{Simulation parameters}  
	\label{table1} 
	\begin{tabular}{|c|c|}
		\hline
		Number of APs $M$ &5\\
		\hline
        Number of UEs $U$ &3\\
		\hline
        Frequency $f_{c}$ & 60GHz\\
        \hline
        Slot Length $\Delta T$ & 50 ms\\
        \hline
        Transmission Frame & 0.5 s\\
        \hline
        Transmission Power of UEs & 23 dBm\\
        \hline
        Bandwidth & 500 MHz\\
        \hline
		Transmission Power of APs $p_{m}$ &23 dBm\\
		\hline
		Number of Transmitting Antennas in AP $N_t$&32\\
        \hline
		Number of Receiving Antennas in AP $N_r$&32\\
        \hline
		Number of Antennas in UE $N_u$ &4\\
        \hline
		ST Moving Speed $v_s$ &2 m/s\\
        \hline
		ST Length $l_s$ &5 m\\
        \hline
		ST Width $w_s$ &2 m\\
        \hline
		Number of Scatters $K$ &8\\
		\hline
		Noise power $\sigma ^2$ &-87 dBm\\
		\hline
		$a_{\tau}$&$6.7*10e^{-7}$\\
        \hline
		$a_{\mu}$&$2*10^{4}$\\
        \hline
		$a_{\theta}$&$1$\\
        \hline
	\end{tabular}
\end{table}

\section{Simulation Results}
In this section, we evaluate the performance of our proposed sensing-assisted channel estimation framework, and compare it against the traditional LS channel estimation method. The indoor scenario is modeled as a 200$m$ square, where the left bottom is the original point, i.e., $\mathrm{O}(0, 0)$. Five APs are deployed at $\mathrm{{AP}_1} (0, 0)$, $\mathrm{{AP}_2} (200, 200)$, $\mathrm{{AP}_3} (0, 200)$, $\mathrm{{AP}_4} (200, 0)$, and $\mathrm{{AP}_5} (100, 200)$ respectively. Three UEs are deployed at $u_1 (50, 150)$, $u_2 (150, 150)$, and $u_3(150, 100)$, respectively. We assume the moving target ST moves towards the x-axis positive direction with the speed of $2\mathrm{m/s}$ starting from $\mathrm{O}_{\mathrm{t}}(0, 50)$. It is noted that the length and width of the moving target are $5\mathrm{m}$ and $2\mathrm{m}$, respectively. Detailed parameters are given in Table.~\ref{table1}.

\begin{figure}[!h]
\centering
\subfigure[Estimated position and ground truth position of ST]
{\includegraphics[scale=0.35]{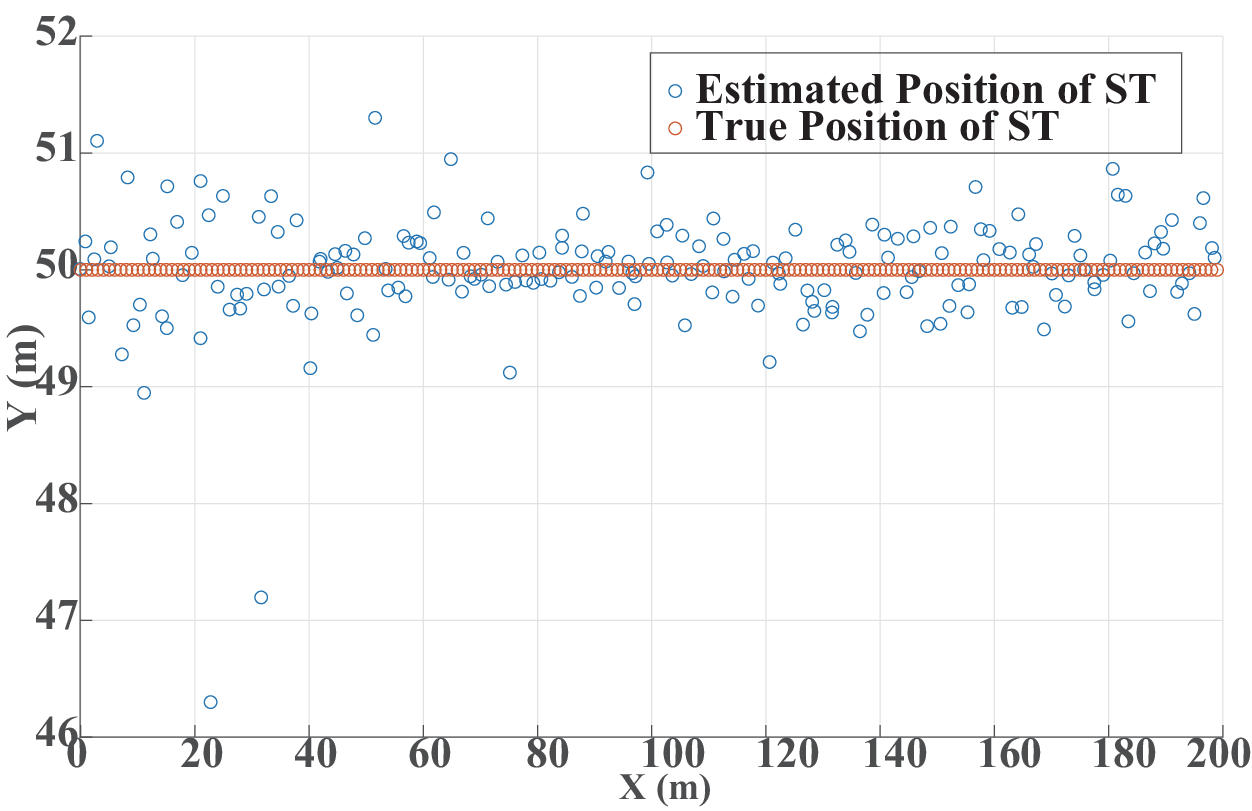}} 
\centering
\subfigure[Correlation between estimated channel and ground truth channel under various number of APs]
{\includegraphics[scale=0.35]{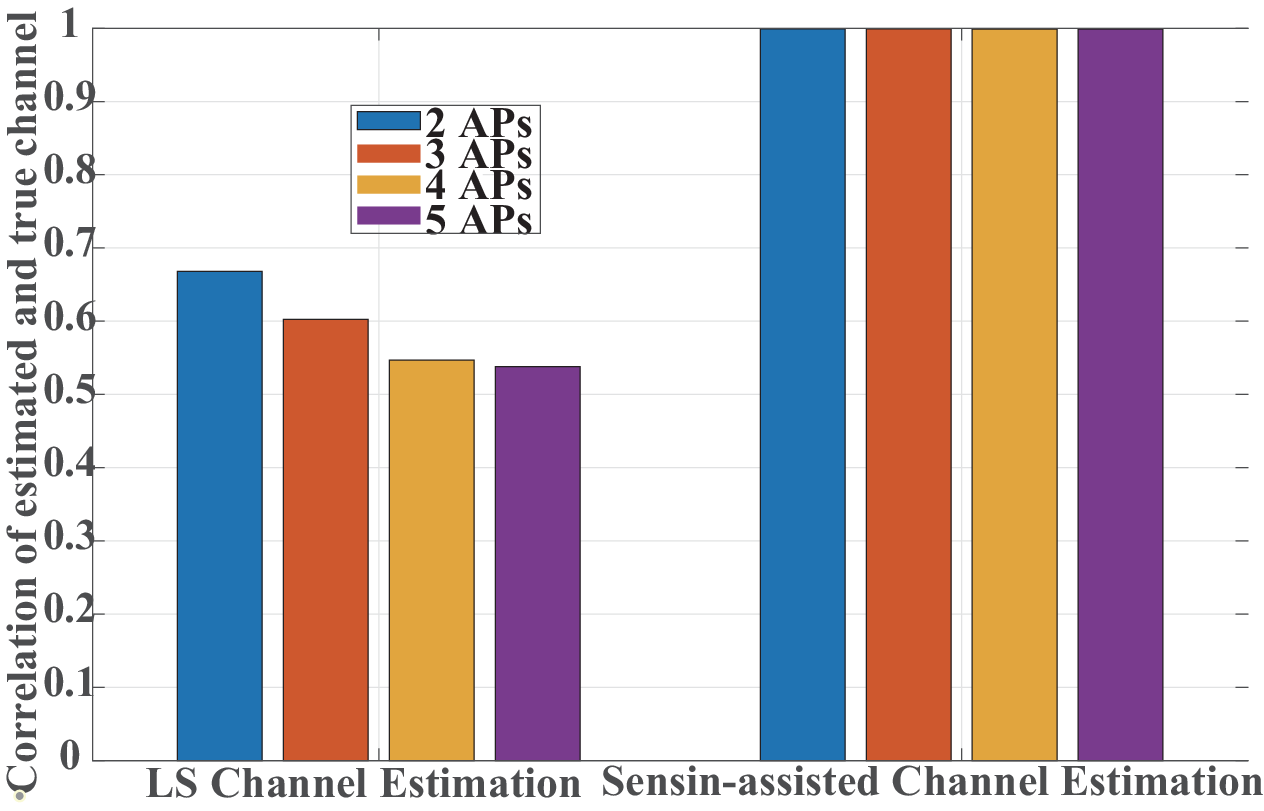}}
\caption{Estimated position of ST and channel correlation coefficient}
\label{fig:throughput}
\end{figure}

Fig.~\ref{fig:throughput} illustrates the estimated positions of ST, and channel correlation coefficient in traditional LS and sensing-assisted channel estimation. In Fig.~\ref{fig:throughput} (a), we can observe that multiple APs successfully estimate and track the position of ST, where the mean estimation error is 0.37m. In Fig.~\ref{fig:throughput} (b), we present the correlation coefficient between the estimated channel and ground truth channel, which is calculated as $\frac{1}{U}\sum_{u \in \mathcal{U}} \frac{trace(\mathbf{{h}_{u}^{j, n}}\mathbf{\widehat{h}_{u}^{j,n\dagger}})}{||\mathbf{\widehat{h}_{u}^{j,n}}||_{\mathrm{F}}||\mathbf{\widehat{h}_{u}^{j,n}}||_{\mathrm{F}}}$ to quantify the correlation between ground truth paths and estimated paths. We can observe that the proposed sensing-assisted channel estimation scheme achieves over 0.99 correlation coefficient between the estimated channel and ground truth channel, where the correlation coefficient of LS channel estimation only achieves 0.6 correlation coefficient and decreases with the increasing number of APs, due to the increase of channel dimension.

\begin{figure}[!h]
\centering
\subfigure[Average throughput under various number of APs]
{\includegraphics[scale=0.35]{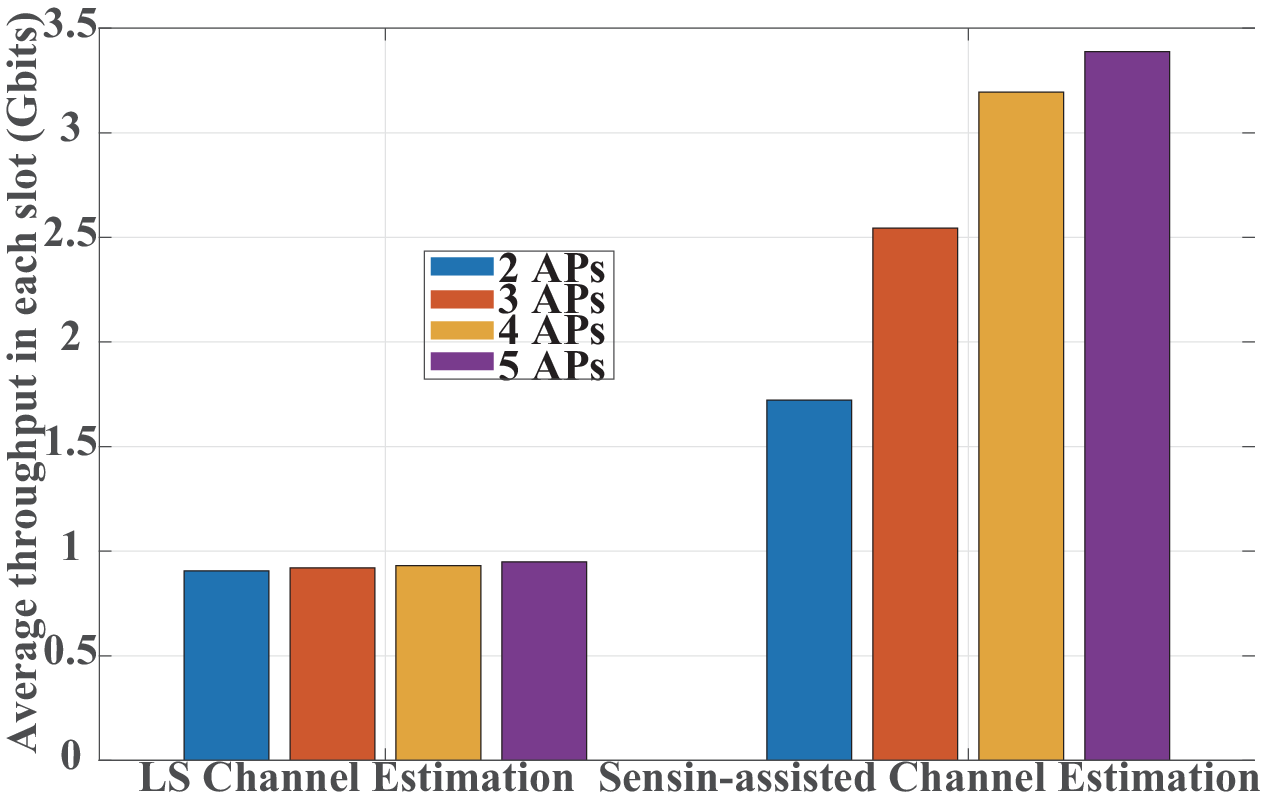}} 
\centering
\subfigure[Average throughput under various transmission power of APs]
{\includegraphics[scale=0.35]{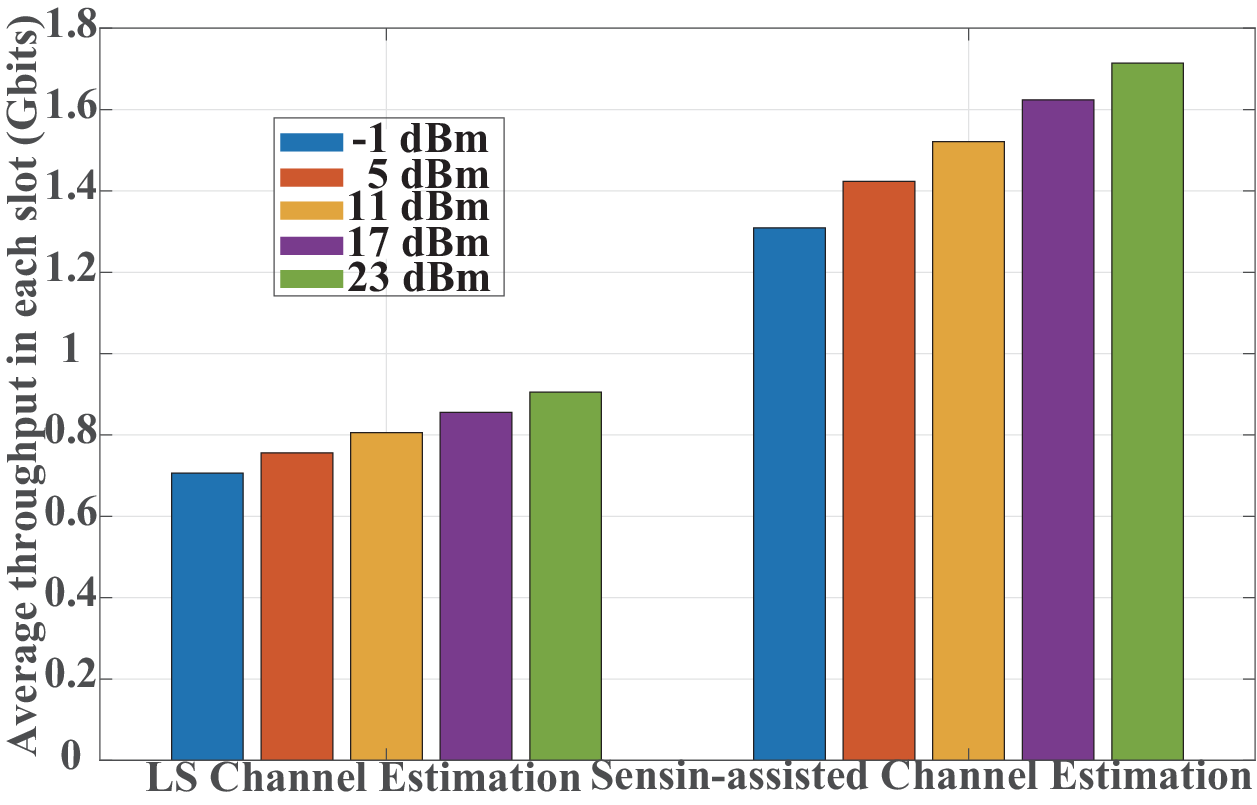}}
\caption{Average throughput under various numbers of APs and transmission power.}
\label{fig:AP_SPEED}
\end{figure}

Fig.~\ref{fig:AP_SPEED} illustrates the average throughput in each slot for varying number of APs and various transmission powers. In Fig.~\ref{fig:AP_SPEED} (a), we can observe that the average throughput increases significantly with the increasing number of APs for the proposed sensing-assisted channel estimation framework. However, increasing the number of APs does not significantly increase the throughput gain in the traditional LS estimation. This is because the DMIMO with a higher number of APs requires stringent channel estimation accuracy for interference mitigation, which cannot be satisfied by the traditional LS estimation. In Fig.~\ref{fig:AP_SPEED} (b), we can observe that the average throughput of both LS estimation and sensing-assisted estimation increases with increasing APs' transmission power. However, this increase is more pronounced with the proposed method.

\section{Conclusion}
In this paper, we proposed a sensing-assisted channel estimation framework, which exploits the sensing capability of multiple APs in a DMIMO network to enhance channel estimation performance via the Ray-tracing method. Different from the traditional sensing-assisted communication works focusing on LoS scenarios, the proposed scheme can accurately estimate the NLoS user channel in a dynamic environment caused by the moving targets, which achieves significant throughput performance gain compared to the traditional LS estimation.
\bibliographystyle{IEEEtran}
\bibliography{IEEEabrv,references}

\end{document}